\documentstyle[prl,aps,preprint,noheadings,epsf]{revtex}

\begin{document}
\tighten
\newcommand{\ttt}{\hat t}
\newcommand{\sss}{\hat s}
\newcommand{\uuu}{\hat u}
\title{QCD SELECTION RULES IN POLARIZED HADRON COLLISIONS\thanks
{This work is supported in part by funds provided by the U.S.
Department of Energy (D.O.E.) under cooperative
agreement \#DF-FC02-94ER40818 and \#DE-FG02-92ER40702
and in
part by funds provided by the National Science Foundation under grant
\# PHY 92-18167.}}

\author{R. L. Jaffe}

\address{Center for Theoretical Physics, Laboratory for Nuclear Science \\
and Department of Physics \\
Massachusetts Institute of Technology, Cambridge, Massachusetts 02139 \\
and\\
Lyman Physics Laboratory, Harvard University\\
Cambridge, Massachusetts 02138 \\
{~}}{\centering and}

\author{N. Saito}

\address{Radiation Laboratory\\
RIKEN (The Institute of Physical and Chemical Research) \\
Wako, Saitama, 351-01, Japan \\
{~}}{}

\date{MIT-CTP-2519 ~~~ HUTP-95/A013 Submitted to
{\it Physics Letters B} ~~~ March 1996}
\maketitle

\begin{abstract}

Plans are underway to measure spin asymmetries at large momentum
transfer in  hadron hadron collisions at RHIC and elsewhere. 
Proposals have focused on  measuring quark transversity and quark and
gluon helicity distributions in  the nucleon.  These experiments will
also provide  a strong and simple test of perturbative QCD, namely
that ${\cal A}_{TT}/{\cal A}_{LL}\ll 1$  in $pp\rightarrow 2\, {\rm
jets} +X$ and several related processes, whereas 
${\cal A}_{TT}/{\cal A}_{LL}\sim 1$ in Drell-Yan  production of muon
pairs.  The prediction  tests the helicity, twist  and chirality
selection rules of perturbative QCD that form  the foundation for the
analysis of spin dependent hard processes.  We estimate the ratio
${\cal A}_{TT}/{\cal A}_{LL}$ numerically for  polarized protons
at RHIC.

\end{abstract}

\break
Gluons are abundant in the nucleon.  In addition their color charge is larger 
than that of quark and antiquarks.  So in general gluon-gluon ($gg$) and 
gluon-quark ($gq$) scattering dominate hard processes unless forbidden by 
initial or final state specifications.
Thus, for example, 
$gg$, $gq$ scattering dominate two jet ($jj$) production in $pp$ collisions 
except at large $x_1$ and $x_2$ where valence quarks overwhelm gluons.
As an example where selection of a final state removes gluons from 
consideration one need look no further than the classic Drell-Yan process 
at leading order in $\alpha_s$,
since production of lepton pairs requires $q\bar q$
annihilation.  As a result $pp\rightarrow \ell\bar\ell X$ is considerably 
suppressed compared to $pp\rightarrow jj  X$ for comparable kinematics
(trivial factors of $\alpha_s$ and $\alpha_{EM}$ aside).

The possibility that {\it polarized\/} gluons are abundant 
in a polarized nucleon is now 
taken quite seriously.  It therefore seems appropriate to look at the
relative magnitude of spin dependent 
hard processes depending whether polarized
gluons can contribute.  The helicity and chirality structure of the 
leading-twist quark and gluon distributions have a dramatic effect on these 
considerations.  The helicity difference gluon distribution, $\Delta g 
(x, Q^2)$, is of considerable interest in connection with the nucleon spin 
problem and may be large.\cite{deltag}  However,
{\it there is no helicity flip gluon distribution 
at leading twist\/} and therefore {\it gluons do not contribute to transverse 
asymmetries at leading twist\/} since transverse asymmetries originate 
from the helicity flip amplitude.  As a result there is a striking difference 
between the transverse asymmetry, ${\cal A}_{TT}$ and the 
longitudinal asymmetry, ${\cal A}_{LL}$ in processes dominated by $gg$ and/or 
$gq$ scattering.  In contrast, processes like Drell-Yan where gluons 
do not contribute, show no dramatic difference between ${\cal A}_{TT}$ and 
${\cal A}_{LL}$.

In this Letter we review the helicity and chirality assignments of 
quark and gluon distributions at leading twist and derive the ``selection 
rules'' that are our principle results.  Next we briefly review earlier work 
and explain what we have added to the subject.  Finally we take some 
reasonable models of polarized quark and gluon distributions in the nucleon 
and use {\sc Pythia}, one of the widely-used QCD event generators,
to make numerical estimates for a future polarized proton-proton 
collider environment at RHIC.

\section{Parton Distribution Functions and Selection Rules in Polarized Hadron 
Scattering}
The helicity, chirality and transversity classification of
parton distributions within a spin-$1/2$ target are now well 
understood through twist-three.\cite{erice}  Dominant effects are 
determined by twist-two distributions which scale (modulo logarithms) in the 
deep inelastic limit.  Subdominant effects, which are suppressed by a
power of a large momentum scale, are described by twist-three distributions.  

The twist-two quark distributions are $q(x,Q^2)$, $\Delta q(x,Q^2)$, and 
$\delta q(x,Q^2)$,
\begin{itemize}
\item $q^a(x,Q^2)$ is the familiar spin averaged quark distribution, which 
preserves quark chirality.  $a$ is a flavor label.
\item $\Delta q^a(x,Q^2)$ is the helicity difference quark distribution.  It 
measures the probability to find a quark with helicity parallel minus 
antiparallel to the target helicity.  $\Delta q$ also preserves quark chirality.
\item $\delta q^a(x,Q^2)$ is the transversity difference 
quark distribution.  It 
measures the probability to find a quark polarized parallel minus 
antiparallel to the polarization of a target polarized transverse to its 
(infinite) momentum.  In a helicity basis, $\delta q^a$ is off-diagonal in the 
helicity of both the quark and target.  The transverse eigenstates can be 
written as superpositions of helicity 
eigenstates  --- $\vert \top,\bot \rangle = 
{1\over\sqrt{2}}(\vert +\rangle \pm \vert -\rangle)$
In the squared amplitude, the difference of $\top\top$ 
and $\top\bot$ isolates helicity flip in the helicity basis.
The quark helicity changes by 
$\Delta\lambda=\pm 1$ and the target helicity changes by $\Delta\Lambda=\mp 
1$.  $\delta q^a$ flips chirality as well as helicity and 
therefore decouples from deep inelastic lepton scattering.
\end{itemize}

The gluon distributions at twist-two have the same structure 
($g(x,Q^2)$ and $\Delta 
g(x,Q^2)$) with the striking exception that there is no gluon distribution
analogous to $\delta q$.  Because the independent gluon 
states have helicity $\lambda=\pm 1$, a helicity flip gluon 
distribution would require the target to absorb $\Delta\Lambda=\mp 2$, which 
is forbidden for a spin-${1\over 2}$ target.  A target with spin greater than 
${1\over 2}$ {\it can\/} have a helicity flip gluon distribution in leading 
twist.  It is known as $\Delta(x,Q^2)$ and some of its properties are explored 
in Refs.~\cite{gt}.  Also, there is a helicity flip gluon 
distribution at twist-three where $\lambda=0$ can appear, but 
its effects are suppressed by a power of a 
large momentum scale in hard processes.  The properties of 
these distributions are summarized in Table I.
\begin{table}
\begin{center}
\begin{tabular}{|l|l|ccc|}
   & Helicity     &  Average &  Difference & Flip \\
\tableline
      & Notation & $q(x,Q^{2})$  & $\Delta q(x,Q^{2})$ 
&  $\delta q(x,Q^{2})$ \\
Quark & Helicity amplitudes 
& $A_{\frac{1}{2}\frac{1}{2},\frac{1}{2}\frac{1}{2}}
                     + A_{\frac{1}{2}-\frac{1}{2},\frac{1}{2}-\frac{1}{2}}$
                     & $A_{\frac{1}{2}\frac{1}{2},\frac{1}{2}\frac{1}{2}}
                     - A_{\frac{1}{2}-\frac{1}{2},\frac{1}{2}-\frac{1}{2}}$
                     & $A_{-\frac{1}{2}\frac{1}{2},\frac{1}{2}-\frac{1}{2}}$ \\
      & Chirality & even  & even  & odd  \\
\tableline
      & Notation & $g(x,Q^{2})$  & $\Delta g(x,Q^{2})$ &  $-$  \\
Gluon & Helicity Amplitudes & $A_{\frac{1}{2}1,\frac{1}{2}1}
                     + A_{\frac{1}{2}-1,\frac{1}{2}-1}$
                     & $A_{\frac{1}{2}1,\frac{1}{2}1}
                     - A_{\frac{1}{2}-1,\frac{1}{2}-1}$
                     &   \\
      & Chirality & even  & even  & $-$ \\
\end{tabular}
\caption{\sf Twist-two quark and gluon distribution in the nucleon classified
according to target helicity and quark helicity. Helicity amplitudes for
virtual quark/nucleon forward scattering are denoted
$A_{\Lambda \lambda, \Lambda' \lambda'}$. Distributions are classified
according to whether quark chirality is conserved (even) or not (odd) in
the quark/nucleon amplitude.}
\end{center}
\label{tbl:nucleon}
\end{table}

As a first example we consider $pp\rightarrow jjX$ at energies and momentum 
transfers in the perturbative QCD regime.  As usual, the cross section can be 
written as the product of parton distribution functions and
parton-parton hard 
scattering cross sections.  The parton distributions must be chosen according 
to the polarization of the initial hadrons.  The longitudinal asymmetry, 
${\cal A}_{LL}$ receives contributions from polarized gluons through $\Delta 
g$, and polarized quarks (or antiquarks) through $\Delta q$ ($\Delta \bar 
q$).  
\begin{equation}
{\cal A}^{jj}_{LL}={[\Delta g\otimes\Delta g]\Delta\sigma^{jj}_{gg} + 
[\Delta g\otimes\Delta q]\Delta\sigma^{jj}_{gq}+
[\Delta q\otimes\Delta q]\Delta\sigma^{jj}_{qq}+ 
[\Delta q\otimes\Delta \bar q]\Delta\sigma^{jj}_{q\bar q}+ 
\ldots\over [g\otimes 
g]\sigma^{jj}_{gg} +[g\otimes q]\sigma^{jj}_{gq} + 
[q\otimes q]\sigma^{jj}_{qq}+ 
[q\otimes \bar q]\sigma^{jj}_{q \bar q}+ \ldots}.
\end{equation}
in an obvious notation.  The terms denoted by $\ldots$ 
include anti-quark gluon and anti-quark anti-quark scattering.

In contrast the transverse asymmetry receives contribution {\it only from 
$qq$, $q\bar q$ and $\bar q \bar q$\/} because there is no helicity flip gluon 
distribution.  Furthermore the requirement of helicity conservation in hard 
scattering requires $qq$ or $\bar q\bar q$ scattering to proceed via an 
interference between direct and exchange graphs as shown in 
Figure~\ref{qqscatt}, reducing the contribution to the cross section by a factor 
of $\sim {1\over 11}$.\cite{HMS,Ji}.  The net result is
\begin{equation}
{\cal A}^{jj}_{TT}={
[\delta q\otimes\delta q]\delta\sigma^{jj}_{qq}+ 
[\delta q\otimes\delta \bar q]\delta\sigma^{jj}_{q\bar q}+
[\delta \bar q\otimes\delta \bar q]\delta\sigma^{jj}_{\bar q\bar q}
\over [g\otimes 
g]\sigma^{jj}_{gg} +[g\otimes q]\sigma^{jj}_{gq} + [q\otimes 
q]\sigma^{jj}_{qq}+ 
[q\otimes \bar q]\sigma^{jj}_{q \bar q}+ \ldots}.
\end{equation}
where the gluons are absent and the $qq$ and $\bar q\bar q$ contributions are 
anomalously small.  So the ratio of transverse to longitudinal 
asymmetries is given by
\begin{equation}
{{\cal A}^{jj}_{TT}\over {\cal A}^{jj}_{LL}}={
[\delta q\otimes\delta q]\delta\sigma^{jj}_{qq}+ 
[\delta \bar q\otimes\delta \bar q]\delta\sigma^{jj}_{\bar q\bar q}+
[\delta q\otimes\delta \bar q]\delta\sigma^{jj}_{q\bar q}\over 
[\Delta g\otimes\Delta g] \Delta\sigma^{jj}_{gg}+ 
[\Delta g\otimes\Delta q]\Delta\sigma^{jj}_{gq}+
[\Delta q\otimes\Delta q]\Delta\sigma^{jj}_{qq}+ 
[\Delta q\otimes\Delta \bar q]\Delta\sigma^{jj}_{q\bar q}+ 
\ldots},
\end{equation}
For convenience the $\sigma^{jj}$,
$\Delta\sigma^{jj}$ and $\delta\sigma^{jj}$ are given in 
Table II.\footnote{Our values for $\sigma$ and $\delta\sigma$ differ slightly
from those quoted by Ji for several 
subprocesses.\cite{Ji}  We are grateful to Dr.~Ji for sharing his
calculations with us.} We found the helicity amplitude formalism of Gastmanns
and Wu particularly well suited to these calculations.\cite{GW}
\begin{table}[ht]
\center{
\begin{tabular}{|c|c|c|c|}
Parton process & Spin Average & Helicity Dependent & Transversity Dependent\\
$ab\rightarrow cd$ & Cross Section --- $\sigma_{ab}^{cd}$ & 
Cross Section --- $\Delta\sigma_{ab}^{cd}$ & Cross Section --- 
$\delta\sigma_{ab}^{cd}$\\
\hline
$q\bar q\rightarrow \gamma^*\rightarrow \ell\bar\ell$ & 
${\uuu^2+\ttt^2\over \sss^2}$ & - ${\uuu^2+\ttt^2\over \sss^2}$ &
${1\over\sss^2}$\\
$q q\rightarrow q q$ & ${\sss^2 + \uuu^2\over \ttt^2} +{\sss^2 + 
\ttt^2\over \uuu^2} - {2\over 3} {\sss^2\over \uuu\ttt}$ &
${\sss^2 - \uuu^2\over \ttt^2} +{\sss^2 -
\ttt^2\over \uuu^2} - {2\over 3} {\sss^2\over \uuu\ttt}$ &
${2\over 3\ttt\uuu}$\\
$q q'\rightarrow q q'$ & 
${\sss^2+\uuu^2\over\ttt^2}$ &
${\sss^2-\uuu^2\over\ttt^2}$ &
--- \\
$q \bar q\rightarrow q \bar q$ & ${\sss^2 + \uuu^2\over \ttt^2} +{\uuu^2 +
\ttt^2\over \sss^2} - {2\over 3} {\uuu^2\over \sss\ttt}$ &
${\sss^2 - \uuu^2\over \ttt^2} -{\uuu^2 +
\ttt^2\over \sss^2} + {2\over 3} {\uuu^2\over \sss\ttt}$ &
${2\over\sss^2}-{2\over 3\sss\ttt}$\\
$q \bar q\rightarrow q' \bar q'$ & ${\uuu^2+\ttt^2\over\sss^2}$ &
$-{\uuu^2+\ttt^2\over\sss^2}$ &
${2\over \sss^2}$\\
$q \bar q'\rightarrow q\bar q'$ & ${\sss^2 + \uuu^2\over \ttt^2}$ &
${\sss^2 - \uuu^2\over \ttt^2}$ & --- \\
$q \bar q\rightarrow gg$ & ${8\over 3}{\ttt^2+\uuu^2\over\ttt\uuu} - 
6{\ttt^2+\uuu^2\over\sss^2}$ & $-{8\over 3}{\ttt^2+\uuu^2\over\ttt\uuu} +
6{\ttt^2+\uuu^2\over\sss^2}$ & ${16\over 3\ttt\uuu} - {12\over\sss^2}$\\
$q \bar q\rightarrow g\gamma$ & ${\ttt^2+\uuu^2\over\ttt\uuu}$ 
& $-{\ttt^2+\uuu^2\over\ttt\uuu}$
& ${2\over\uuu\ttt}$\\
\end{tabular}
}
\caption{\sf Parton cross sections and asymmetries.  Each entry multiplies a 
factor of ${4\pi\alpha_s^2\over9\sss^2}$, except for the first (Drell-Yan) 
which multiplies ${4\pi\alpha_{em}^2e_q^2\over 3\sss^2}$, and the last (direct 
$\gamma$ production) which multiplies ${8\pi\alpha_s\alpha_{EM}\over9\sss^2}$.
In addition, helicity entries multiply $\pm 1$ according to whether the
beam helicities are equal ($+1$) or opposite ($-1$), and
transversity entries 
multiply the kinematic factor $\{\uuu\ttt S_a\cdot S_b-\sss\left( S_a\cdot 
k_a S_b\cdot k_b + S_b\cdot k_a S_a\cdot k_b\right)\}$,
which is proportional 
to $\sin^2\theta\cos 2\phi$ in the parton-parton center of mass frame.}
\label{tbl:partons}
\end{table}

The same analysis applies to many hard processes.  Another experimentally 
important example is high $p_T$ direct photon production, which 
proceeds via $qg\rightarrow q\gamma$, $\bar 
qg\rightarrow \bar q\gamma$, and $q\bar q\rightarrow g\gamma$ at lowest order 
in QCD.  For longitudinal asymmetries all three processes 
contribute, 
but only the last survives for transverse asymmetries.  The ratio of 
asymmetries is therefore
\begin{equation}
{{\cal A}^{\gamma j}_{TT}\over {\cal A}^{\gamma j}_{LL}}={
[\delta q\otimes\delta \bar q]\delta\sigma^{\gamma j}_{q\bar q}\over 
[\Delta g\otimes\Delta q]\Delta\sigma^{\gamma j}_{gq}+
[\Delta g\otimes\Delta \bar q]\Delta\sigma^{\gamma j}_{g\bar q}+
[\Delta q\otimes\Delta \bar q]\Delta\sigma^{\gamma j}_{q\bar q}
},
\end{equation}
The relevant parton cross sections are also given in Table II.
Other hard processes for which selection rule ${\cal A}_{TT}/{\cal A}_{LL}\ll
1$ holds include 
\begin{itemize}
	\item $\vec p \vec p\rightarrow \pi + X$ --- Inclusive production 
	of a pion or other hadron at large transverse momentum requires inclusion
	of a $q\rightarrow\pi$ fragmentation function, which does not affect
	the argument given for two jet production.
	\item $\vec p \vec p\rightarrow Q \bar Q$ --- Open heavy quark 
	production is dominated by $gg$ fusion and therefore exhibits the
	same phenomena as $jj$ production.
\end{itemize}

Only Drell Yan production of charged lepton
pairs through $q\bar q\rightarrow (Z^0, \gamma)\rightarrow\ell\bar\ell$ stands 
out as a case where transverse and longitudinal asymmetries are comparable.  
The ratio is given in terms of the ratio of quark transversity ($\delta q$) 
and helicity ($\Delta q$) distributions,
\begin{equation}
	{{\cal A}^{\ell\bar\ell}_{TT}\over {\cal A}^{\ell\bar\ell}_{LL}}
	=-{\sin^2\theta\cos 2\phi\over 1+\cos^2\theta}\,\,
	{{\sum_a e_a^2\delta 
	q^a(x_1)\delta\bar q^a(x_2) + (1\leftrightarrow 2)}
	\over{\sum_a e_a^2\Delta q^a(x_1)\Delta\bar q^a(x_2) 
	+ (1\leftrightarrow 2)}}.
\label{DY}
\end{equation}
for $q\bar q\rightarrow\gamma^{*}\rightarrow\ell\bar\ell$.
In contrast, Drell-Yan production of $\ell\bar\nu$ via $q\bar q\rightarrow 
W^\pm\rightarrow\ell\bar\nu$ give no transverse asymmetry at leading twist.  
Only left handed quarks and antiquarks couple to $W^\pm$ so no helicity flip 
distribution functions participate at all.  So for these processes 
\begin{equation}
	{{\cal A}^{\ell\bar\nu}_{TT}\over {\cal A}^{\ell\bar\nu}_{LL}}=0,
\end{equation}
at leading twist to all orders in $\alpha_{QCD}$.  The first non-vanishing 
contribution appears at twist-four where $g_T\times \bar g_T$ can yield a 
chirality-even transverse asymmetry.  Finally, before taking eq.~(\ref{DY}) 
too seriously it is important to compute higher order corrections.  If 
polarized gluons in the proton are copious then QCD corrections coming from 
processes like $gq\rightarrow q\gamma^*\rightarrow j\ell\bar\ell$ will 
contribute significantly to 
${\cal A}^{\ell\bar\ell}_{LL}$ but not to ${\cal A}^{\ell\bar\ell}_{TT}$.

\section{Relation to Earlier Work}
The first suggestion that ${{\cal A}^{jj}_{TT}\over
{\cal A}^{jj}_{LL}}$ should be small in QCD was made in a 1979 paper by 
Hidaka, Monsay and Sivers.\cite{HMS}  Their conclusions were quoted 
extensively in the 
1983 review by Craigie, Hidaka, Jacob and Renard.\cite{CHJR}  Their work is 
handicapped by the incomplete understanding of transverse spin in hard 
processes available at that time.  In particular, gluons were ignored entirely 
and quark transversity distributions were described by the twist-three 
distribution function $g_2$.
A $qg$ contribution to ${\cal A}_{TT}$ appears in
the unnumbered equation following eq.~(8) in Ref.~\cite{HMS} but is 
subsequently dropped without mention.  Their
treatment of transversely polarized 
quarks is incorrect.  If transversely polarized quarks were described by 
$g_2$, then their effects would be suppressed in $q\bar q\rightarrow jj$ by a 
factor of $1/s$ since $g_T$ is twist-three.  Ref.~\cite{HMS} mistakenly 
treated the Wandzura-Wilczek contribution to $g_T$ as if it scaled.\cite{WW}
Despite the errors in their argument, the conclusion that ${{\cal 
A}^{jj}_{TT}\over{\cal A}^{jj}_{LL}}$ is very small remains correct.

The publication of Ralston and Soper's 1979 paper introducing the 
twist-two transversity distribution made it possible to reconsider 
${{\cal A}^{jj}_{TT}\over{\cal A}^{jj}_{LL}}$ in the manner we have 
presented.\cite{RS}  Recently many authors have considered transverse 
asymmetries in production and fragmentation processes at high momentum 
transfer.\cite{Ji,erice}  The focus was on developing the proper
interpretation  of transverse polarization phenomena, on the
information provided by measurements of transverse asymmetries and on the 
description of processes where the effects could be measured.
Only Ji noted that transverse asymmetries in jet production 
are typically small and do not receive $gg$ contributions.\cite{Ji}  
None of the papers pointed out the basis of the 
selection rule ${\cal A}_{TT}/{\cal A}_{LL}\ll 1$.  Our purpose is to call 
attention to the fundamental importance of this result and to provide some 
numerical estimates of its magnitude.

\section {Estimates for a Polarized Collider}

The first collision of polarized protons at RHIC is scheduled for 1999.
The beam energy will be variable and the center-of-mass energy will range
$50 < \sqrt{s} < 500$~GeV.  The expected luminosity is
$2.0~(0.8)\times 10 ^{32}$cm$^{-2}$sec$^{-1}$ at $\sqrt{s}=$500~(200)~GeV.
The polarization of the proton beam is expected to be 70\%.
The most plausible integrated luminosity is 800~(320)~pb$^{-1}$ at
$\sqrt{s}=$500~(200)~GeV per year. In this Letter, we focus on the
$pp$ collisions at $\sqrt{s}=500$~GeV. 

We have utilized the QCD event generator {\sc Pythia}~\cite{PYTHIA}
for numerical studies of the spin asymmetries.
We have calculated the
asymmetries, ${\cal A}_{LL}$ and ${\cal A}_{TT}$, in various QCD
processes with the polarized structure function sets 
obtained by Gehrmann and Stirling (GS95)~\cite{GS95} through
next-to-leading order analysis\footnote{Their model includes
the set obtained through leading order analysis. In our analysis
only next-to-leading order version is employed.} of the experimental data
and the asymmetries calculated from the spin-dependent
and spin-independent cross sections listed in Table~\ref{tbl:partons}.

The polarized structure function set GS95 has three models of
gluon polarization, all of which give equivalently good description of
existing
experimental data. The models GS95(A), (B), and (C) corresponds to
large, medium, and small gluon polarization, respectively.

In our calculation, we have assumed that the transversity difference quark
distribution $\delta q(x,Q^{2})$ is identical to the helicity
difference quark distribution $\Delta q(x,Q^{2})$. There are
several reasons for this choice.  First, there is no data on $\delta q$ to 
guide us.  Second, models typically find $\delta q\sim \Delta q$.
The bag model predicts
$\delta q$ of a shape and magnitude very similar to 
$\Delta q$;\cite{JAFFEJI91}
a calculation by means of QCD sum rules suggests $\delta q$ which
is smaller than $\Delta q$, but not by order of magnitude;\cite{QCDSR} and 
preliminary results on the lattice calculation~\cite{HATSUDA} of 
{\it tensor charge}
which is the integrated value of quark transversity distributions,
suggest $\delta \Sigma =0.64 \pm 0.12.$  It should be noted that 
$Q^{2}$-evolution of $\delta q$ is
completely different from  $\Delta q$ so the assumption we have made cannot 
be valid for entire $Q^{2}$ region.  In this situation the simplest
assumption will allow further modification of the presented
results rather easily when our knowledge of $\delta q$ improves.

\subsection{Inclusive Jet Production}

All quark and gluon related tree level processes in the QCD have
been simulated with {\sc Pythia}. 
In the low $p_{T}$ region,
gluon-gluon scattering dominates the cross section, and quark-quark
contribution is dominant at high-$p_T$ region.
Annihilation channels contribute only a very small fraction of order 
10$^{-3}$. 
No rapidity cut has been applied to the final jet in this calculation.

The asymmetries ${\cal A}_{LL}$ and ${\cal A}_{TT}$
for inclusive jet production have been calculated with
GS95-models, (A), (B), and (C). The results are shown
in Figure~\ref{F:ASYM_JET}.
The values of ${\cal A}_{TT}$ at $\phi=0$ are plotted for
a comparison with ${\cal A}_{LL}$.
The size of transverse asymmetry ${\cal A}_{TT}$ is
about $5\times 10^{-3}$ at $p_T$=100~GeV/$c$. Such a small asymmetry
will be hard to detect with RHIC detector systems with assumed integrated
luminosity, so for practical purposes the QCD prediction is the ``selection 
rule'' ${\cal A}_{TT}/{\cal A}_{LL} = 0$.

Clearly the longitudinal asymmetry  ${\cal A}_{LL}$
for inclusive jet production is quite sensitive to the
abundance of polarized gluons while ${\cal A}_{TT}$ is insensitive.
We calculated $\frac{{\cal A}_{TT}}{{\cal A}_{LL}}$ 
for each $\Delta g$ model, and show the results 
in Figure~\ref{F:RATIO_JET}
excepting the GS95(C) model, which gives an infinite value due to the
zero-crossing in ${\cal A}_{LL}$.

It might be surprising that
our $A_{LL}$ values are rather smaller than previous
calculations~\cite{SOFFER90}. Especially $qq$-scattering, which is the
dominant process at high-$p_T$, contributes very little to the asymmetry.
This is due to the cancellation of the asymmetry 
for $uu$-scattering by the asymmetry for $ud$-scattering, since
$\frac{\Delta d}{d}$ has opposite sign to
$\frac{\Delta u}{u}$ and the partonic level asymmetry
is larger for the scattering of different flavors than
for the scattering of identical flavor. 
The strength of this cancellation could change significantly, if
we take the model by Bourrely and Soffer~\cite{BS}, where flavor
decomposition of polarized quark distribution is quite different from 
the GS models.
This cancellation does not
work on the ${\cal A}_{TT}$ since only the scattering of identical
flavors contribute to the asymmetry. 

\subsection{High-$p_{T}$ Direct Photon Production}

The transverse and longitudinal asymmetries for
high-$p_T$ direct photon production are calculated in a similar way
to the inclusive jet production.
We have applied a cut on pseudo-rapidity, $|\eta|<0.35$ which 
is the acceptance of the electromagnetic calorimeter of PHENIX detector 
system at RHIC. 
The results are shown in Figure~\ref{F:ASYM_DG}
Again, values at $\phi=0$ are plotted for
${\cal A}_{TT}$ to compare with ${\cal A}_{LL}$.
The longitudinal asymmetries are order of a few percent,
and are dependent on the assumed gluon polarization. 
The transverse asymmetries are at the 10$^{-3}$ level. 
The ratio $\frac{{\cal A}_{TT}}{{\cal A}_{LL}}$ is calculated
for each $\Delta g$ model, and plotted in Figure~\ref{F:RATIO_DG}.
The ratio ranges within 20\% in the $p_T$ region where we expect
significant statistics ($<30$~GeV/$c$)
and again will be hard to detect
experimentally at RHIC.

We conclude that the
predicted transverse asymmetries in two jet and direct photon
production are very small and, within the sensitivity of 
planned experiments, 
consistant with zero.  Confirmation of ${\cal A}_{TT}/{\cal A}_{LL}\ll 1$ 
tests fundamental aspects of spin structure in perturbative QCD.
Any nonzero asymmetry for the processes
studied above, if observed,  would require reconsideration of
all spin effects in perturbative QCD.

\section{Acknowledgements}

One of us (RLJ) would like to thank Xiangdong Ji for discussions, 
and the Physics Department at Harvard University for hospitality.  
We are grateful to Thomas Gehrmann for providing us a FORTRAN
code to calculate the polarized structure functions. 
We would 
like to thank the organizers of the Adriatico Conference on Trends in Collider 
Spin Physics and the RIKEN Symposium on Spin Structure of the Nucleon where 
this collaboration was begun. 
A part of this work has been done within the framework of the RIKEN-BNL
collaboration for the RHIC/Spin project.

\bibliographystyle{unsrt}

\begin{center} 
Figures

\begin{figure} 
\epsfxsize=5in
\centerline{\epsffile[61 279 553 514]{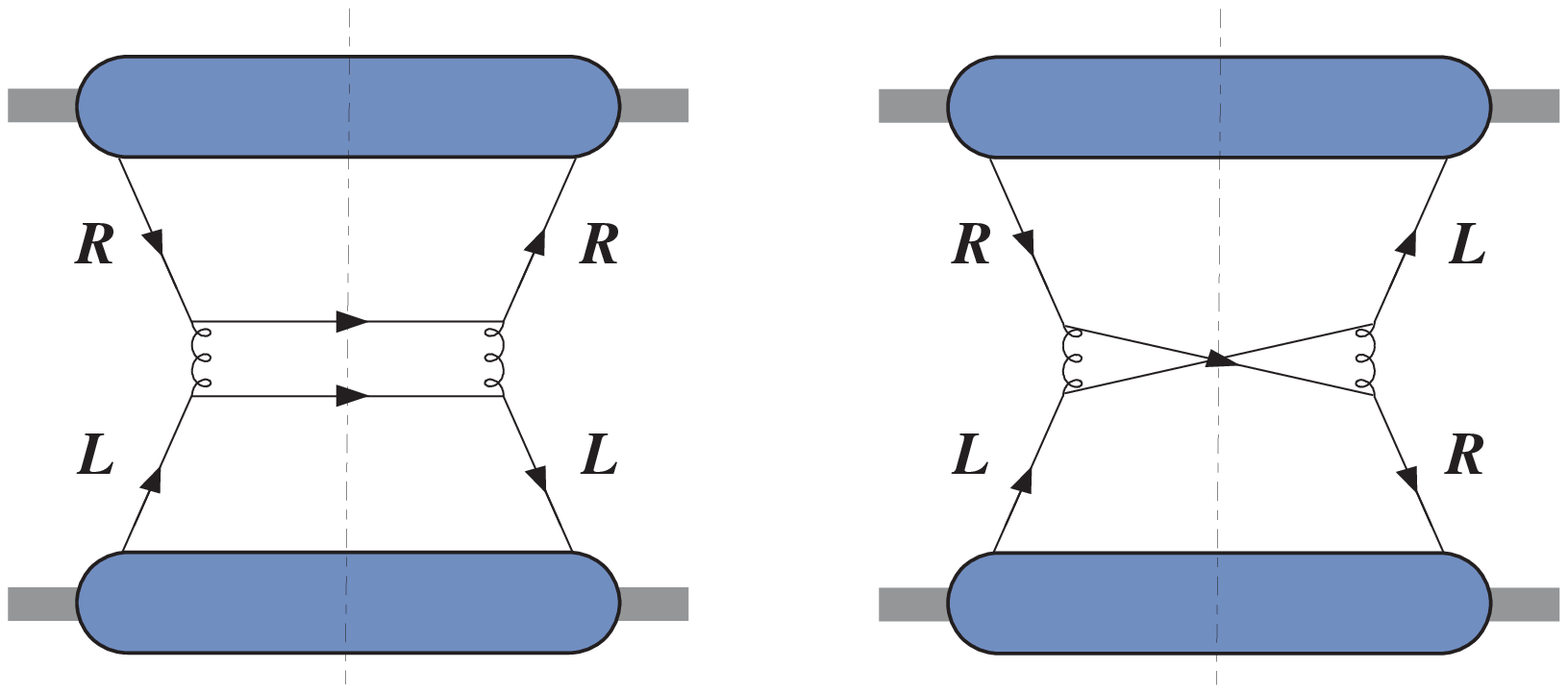}}
\caption{Direct and Exchange graphs for $qq$ or $\bar{q}\bar{q}$ 
scattering.}
\label{qqscatt}
\end{figure} 

\vskip.3in

\begin{figure} 
\epsfxsize=5in
\centerline{\epsffile[71 181 524 542]{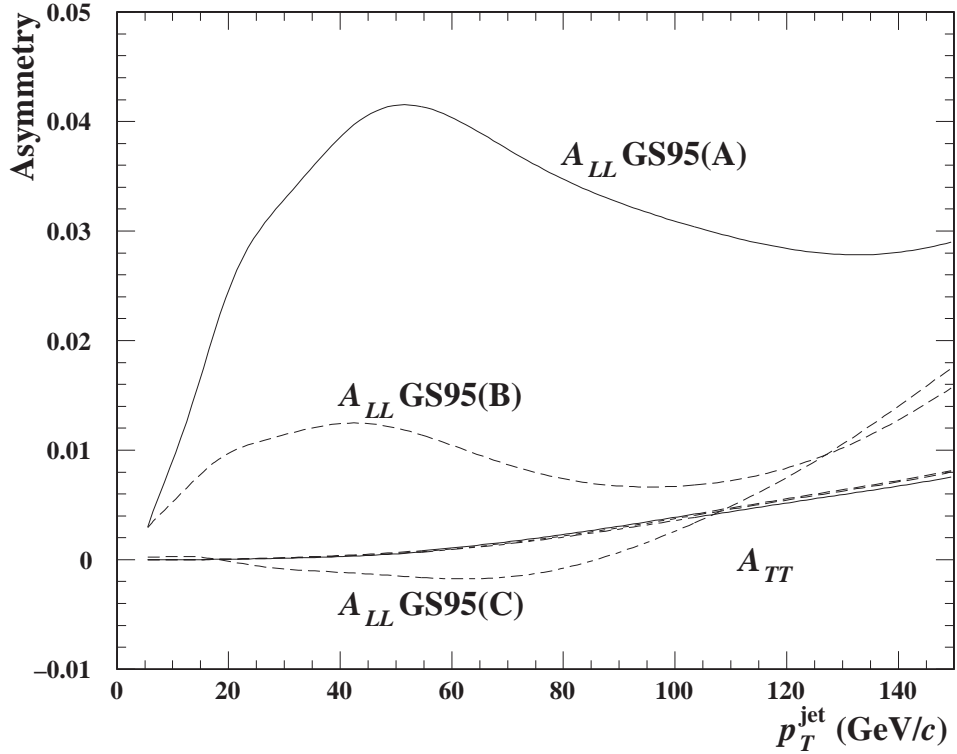}}
\caption{Longitudinal and transverse asymmetries for inclusive 
jet production in $pp$-collisions at $\protect\sqrt{s}=$500~GeV.
Asymmetries are calculated with three models of gluon polarizations,
GS95(A) (solid), GS95(B) (dotted), and GS95(C) (dashed-dotted). }
\label{F:ASYM_JET}
\end{figure}

\begin{figure}
\epsfxsize=5in     
\centerline{\epsffile[71 181 524 542]{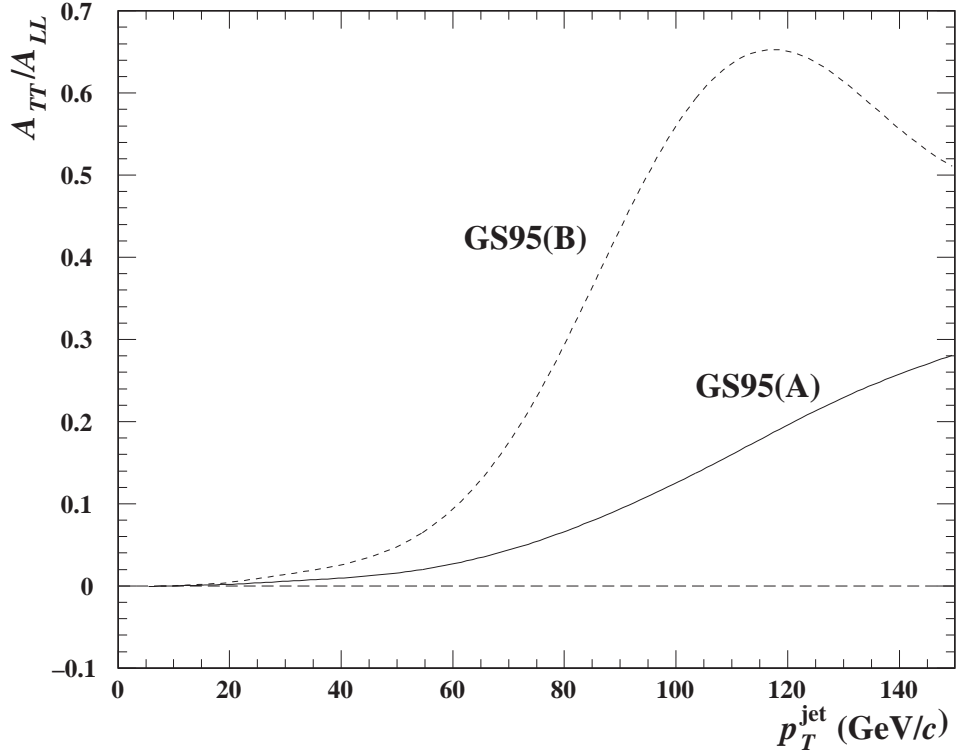}}
\caption{Ratios of transverse asymmetries to longitudinal asymmetries for
inclusive jet production in $pp$ collisions at $\protect\sqrt{s}=$500~GeV.}
\label{F:RATIO_JET}
\end{figure}

\begin{figure} 
\epsfxsize=5in    
\centerline{\epsffile[64 180 530 542]{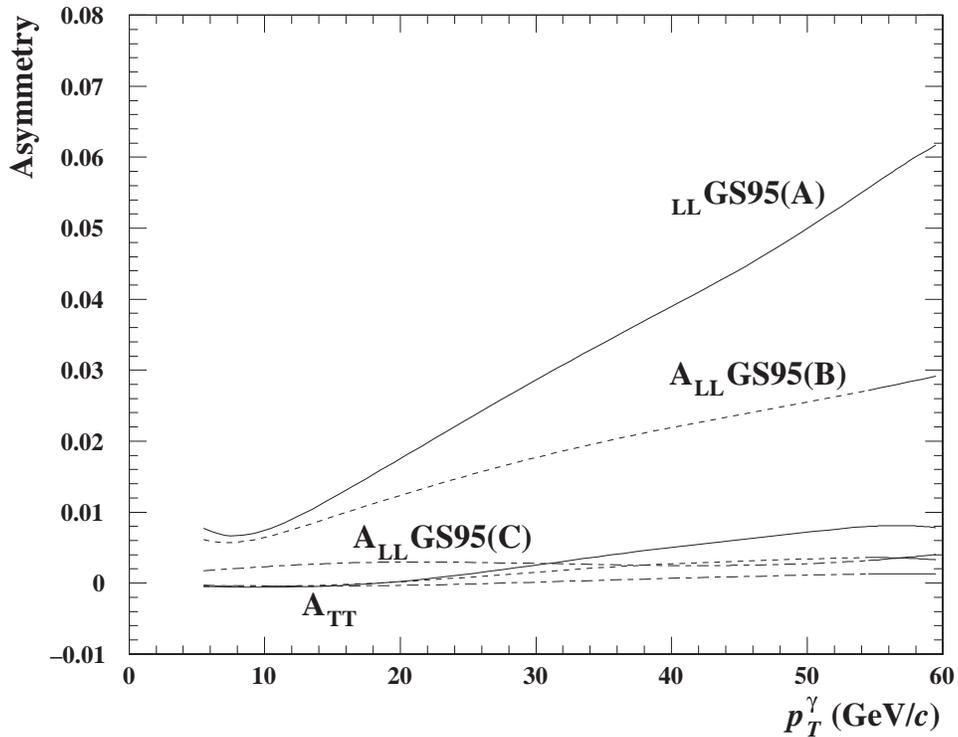}}
\caption{Longitudinal and transverse asymmetries for high-$p_T$
direct photon production in $pp$-collisions at $\protect\sqrt{s}=$500~GeV.
Asymmetries are calculated with three models of gluon polarizations,
GS95(A) (solid), GS95(B) (dotted), and GS95(C) (dashed-dotted).}
\label{F:ASYM_DG}
\end{figure}

\begin{figure} 
\epsfxsize=5in    
\centerline{\epsffile[71 181 530 542]{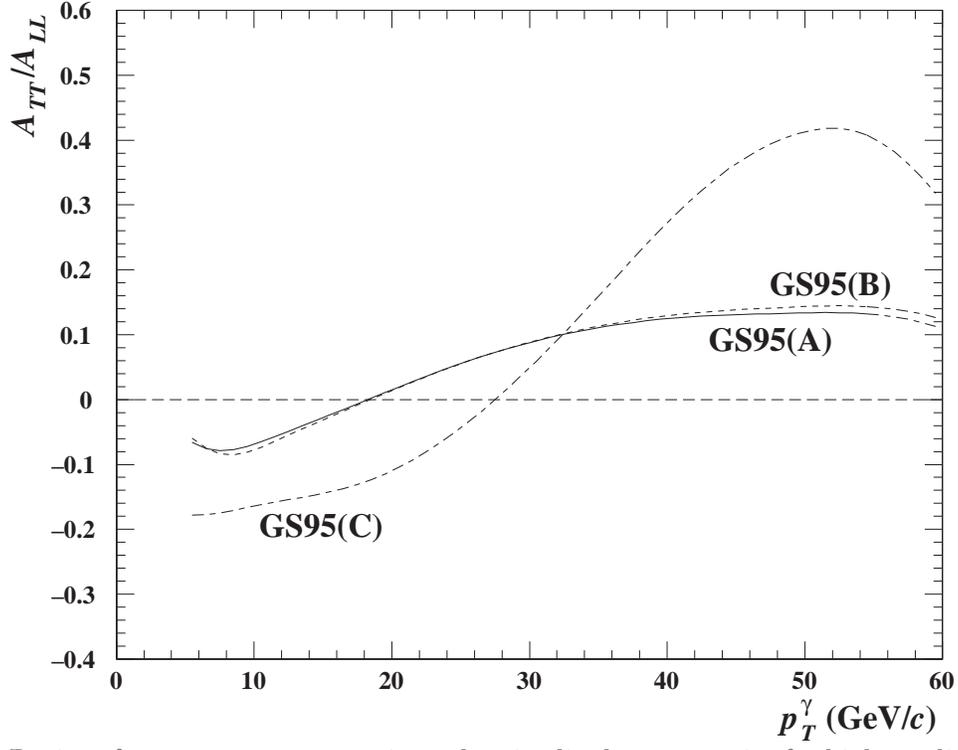}}
\caption{Ratios of transverse asymmetries to longitudinal asymmetries for
high-$p_T$ direct photon production in $pp$ collisions at 
$\protect\sqrt{s}=$500~GeV.}
\label{F:RATIO_DG}
\end{figure} 

\end{center}

\end{document}